\documentclass{article}
\usepackage{ijcai24}

\usepackage{times}
\usepackage{soul}
\usepackage{url}
\usepackage[hidelinks]{hyperref}
\usepackage[utf8]{inputenc}
\usepackage[small]{caption}
\usepackage{graphicx}
\usepackage{amsmath}
\usepackage{amsthm}
\usepackage{booktabs}
\usepackage{algorithm}
\usepackage{algorithmic}
\usepackage[switch]{lineno}
\usepackage{amssymb}
\usepackage{xcolor}
\usepackage{svg}

\usepackage{subcaption}

\title{KnowledgeHub: An End-to-End Tool for Assisted Scientific Discovery}

\author{
Shinnosuke Tanaka$^1$\and
James Barry$^1$\and
Vishnudev Kuruvanthodi$^1$\and
Movina Moses$^2$\and \\
Maxwell J. Giammona$^2$\and
Nathan Herr$^3$\footnote{Work done while at IBM.}\and
Mohab Elkaref$^1$\And
Geeth De Mel$^1$\\
\affiliations
$^1$IBM Research Europe\\
$^2$IBM Research\\
$^3$University College London
\emails
\{shinnosuke.tanaka, james.barry, vishnudev.k, movina.moses, maxwell.giammona, mohab.elkaref\}@ibm.com,
uceenhe@ucl.ac.uk,
geeth.demel@uk.ibm.com
}

\begin{document}

\maketitle

\begin{abstract}
    This paper describes the KnowledgeHub tool,
    a scientific literature Information Extraction (IE) and Question Answering (QA) pipeline.
    This is achieved by supporting the ingestion of PDF documents
    that are converted
    to text and structured representations.
    An ontology can then be constructed where a user defines the types of entities and relationships they want to capture.
    A browser-based annotation tool enables annotating the contents of the PDF documents according to the ontology.
    Named Entity Recognition (NER) and Relation Classification (RC) models
    can be trained on the resulting annotations and
    can be used to annotate the unannotated portion of the documents.
    A knowledge graph is constructed from these entity and relation triples
    which can be queried to obtain insights from the data.
    Furthermore,
    we integrate a suite of Large Language Models (LLMs)
    that can be used for QA and summarisation that is grounded in the included documents via a retrieval component.
    KnowledgeHub is a unique tool that supports annotation, IE and QA,
    which gives the user full insight into the knowledge discovery pipeline.
\end{abstract}

\section{Introduction}

Accelerating scientific discovery has long been the goal of researchers and subject matter experts (SMEs) alike.
The growing amount of data contained within the scientific literature means that automated solutions
are becoming increasingly necessary
to efficiently extract information to develop new discoveries.
Advances in Artificial Intelligence (AI) and Natural Language Processing (NLP) research have led to larger and more capable models
such as BERT \cite{devlin-etal-2019-bert}
that can be used as
feature extractors for token-classification tasks
that have been the cornerstone for Information Extraction (IE) tasks such as Named Entity Recognition (NER) and Relation Classification (RC) between entities.
Additionally,
developments in Large Language Models (LLMs) that predict tokens in an autoregressive manner
\cite{gpt2,NEURIPS2020_1457c0d6,touvron2023llama} and innovative methods like Retrieval Augmented Generation (RAG) \cite{10.5555/3495724.3496517} have led to systems that can leverage vast amounts of internal and external knowledge to enhance
the suitability and factual correctness of LLM responses.

Several annotation tools have been created
that focus on PDF layout annotation,
linguistic annotation
or IE through
NER and RC.
There also exist tools that perform QA over the documents.
In the context of the above,
tools such as PAWLS \cite{neumann-etal-2021-pawls} enable users to annotate PDF document regions with labelled bounding boxes where the tool will then predict layout regions on other files.
Linguistic annotation tools such as PDFAnno \cite{shindo-etal-2018-pdfanno},
AnnIE \cite{friedrich-etal-2022-annie}
and Autodive \cite{du-etal-2023-autodive}
focus on annotating data for
IE tasks such as NER and RC.
Similarly,
BatteryDataExtractor \cite{D2SC04322J} is a tool that focuses on IE for the battery domain with an additional QA component but does not support annotation.

\begin{figure*}[htbp]
    \centering
    \includegraphics[width=0.63\textwidth]{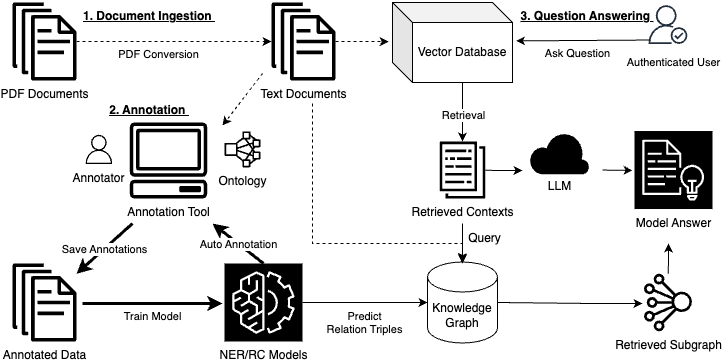} 
    \caption{KnowledgeHub pipeline overview.}
    \label{fig:kh-diagram-full}
\end{figure*}

This paper introduces KnowledgeHub\footnote{A video describing KnowledgeHub is available at this \href{https://youtu.be/TqeRE795v8w}{\bf link}},
a novel tool that
covers the fundamental aspects of the knowledge discovery process:
including
linguistic annotation,
IE with NER and RC models,
and QA that is grounded in the source literature.
To achieve this,
we implement a pipeline where
a user submits a collection of PDF documents for their field of study,
which are then converted to text and structured representations.
A user-defined ontology can then be supplied
which defines the types of entities and relationships to consider.
A browser-based annotation tool enables annotating the contents of the PDF documents according to the ontology.
NER and RC models can then be trained on the resulting annotations
where the trained models can be used to automatically annotate the portion of unannotated documents.
A knowledge graph is constructed from these entity and relation triples which can be queried to gain certain insights.
Furthermore, we include an RAG based QA system.
Out of the systems introduced so far, KnowledgeHub is the only tool that supports
annotation, IE, and QA
(see Table~\ref{tab:comparison} for an overview of different tools).

\section{System Description}

This section describes the KnowledgeHub pipeline.
The overall system is shown in Figure~\ref{fig:kh-diagram-full}.
At a high level,
it is an application that consists of a
frontend built using JavaScript, React and the Carbon Design System library\footnote{\url{https://carbondesignsystem.com/}}.
The frontend is connected to a backend using a Python Flask\footnote{\url{https://github.com/pallets/flask}} web app.
The backend consists of SQLite databases for storing information necessary for running the application.
There is a Neo4j\footnote{\url{https://neo4j.com/}} graph database
as well as a vector store
which are used for building/querying knowledge graphs
and performing RAG.
All the components can be run locally except for the LLM in the RAG pipeline, which requires API calls. We also support a version that is hosted on OpenShift.

Firstly, a user defines a project and specifies the group members and the privileges each member should have.
For that particular project, the remaining pipeline is discussed in the following sections.

\begin{table}[ht]
\centering \small
\begin{tabular}{llc}
\toprule
 \textbf{Tool Name} & \textbf{IE} & \textbf{QA} \\
\midrule
AnnIE & NER, POS & -  \\
Autodive$^\ast$   & NER & - \\
BatteryDataExtractor   & NER, POS, AD  & Extractive \\
PAWLS$^\ast$  & - & -  \\
PDFAnno$^\ast$  & -  & -  \\
\textbf{KnowledgeHub}   & NER, RC & LLM+RAG+KG  \\
\bottomrule
\end{tabular}
\caption{
Comparison with other tools.
\textit{Tool Name}: Superscript($^\ast$) represents support of direct PDF annotation.
AD - Abbreviation Detection;
Extractive - model predicts the answer span;
LLM - model predicts answer with autoregressive LLM.
}
\label{tab:comparison}
\end{table}

\subsection{Document Ingestion}
The user uploads an individual PDF file,
or a zip file containing multiple PDF files.
The PDF content is converted to structured XML using the GROBID tool\footnote{\url{https://github.com/kermitt2/grobid}},
which predicts layout sections
such as
the title, headings, paragraphs, footnotes, references, tables and figures.
We extract the text content inside the paragraph objects predicted by GROBID,
which is then
segmented, tokenised and annotated for Part-of-Speech (POS) information
using the Stanza library \cite{qi-etal-2020-stanza}.

\label{sec:neo4j}

For our setting,
each document is
composed of paragraphs,
and the text within the paragraphs is segmented into sentences.
We create a Neo4j graph database where nodes are created
at the document, paragraph, and sentence level.
The sentence nodes are linked to their origin paragraph node, which in turn is linked to its origin document node.
Metadata such as
the title, authors, and year of publish can be stored for each document.
As we will discuss in Section~\ref{sec:osa}, NER and RC models are used to
predict named entities and relations,
where we also create nodes for the entity tokens and link them with the predicted relation.

The extracted text is also stored in a vector database,
where we use a Chroma\footnote{\url{https://github.com/chroma-core/chroma}} database.
We access an embedding model from the LangChain\footnote{\url{https://github.com/langchain-ai/langchain}} library which offers several embedding models.
Specifically, we use the \texttt{all-mpnet-base-v2} model released by the SentenceTransformers library \cite{reimers-2019-sentence-bert} but note that this model can be easily replaced by a user to suit their needs by changing the model name or embedding provider.

\subsection{Annotation}

An ontology can be created manually in the browser interface or by importing external ontologies such as EMMO\footnote{\url{https://github.com/emmo-repo/EMMO}}.
When importing an ontology, the user selects the entities and their associated relations.
Similarly, our browser-based ontology creation tool enables
i) choosing the entities,
and ii) specifying the relation between certain entities.
This produces a configuration file
which lists the possible relation triples:
\texttt{(entity1, relation, entity2)}.

The user can map these entities and relations
onto the contents of the PDF in a web browser using the BRAT annotation tool \cite{stenetorp-etal-2012-brat}.

We implement our own models for NER and RC.
These models are written in PyTorch \cite{10.5555/3454287.3455008} and involve placing a linear layer on top of a BERT-style model,
where a user can specify an encoding model from the HuggingFace library \cite{wolf-etal-2020-transformers}.
The NER model involves a two-stage process:
first, a span-based model predicts entity span regions including nested structures \cite{yu-etal-2020-named},
then an entity classifier model predicts the entity tag for the selected span \cite{elkaref-etal-2023-nlpeople}.
The RC model predicts a relation type (including no relation) between all pairs of entities in the sentence.
The predictions of the NER/RC models are used to create connected entity nodes in the KG.
A strength of using custom models is that we do not rely on external pipelines such as spaCy and we can train on any data where we have annotations.

\label{sec:osa}
We support two modes of auto-annotation: the first is based on regular expressions to label the target text based on entity names and their types defined by the user.
The second is
machine learning annotation,
where the annotations from BRAT are saved to JSON
and are then used to train NER and RC models.
The trained models are then applied to
the unlabelled data.
This significantly reduces the user burden compared to manual annotation.

\subsection{Question Answering}

RAG is a method used to guide the generation process of an LLM
by providing it with context-appropriate information,
based on retrieving contexts that are most relevant to a user query, e.g. based on the cosine distance between an encoded query and the encoded documents \cite{lewis2021retrievalaugmented}.
KnowledgeHub lets users select a project and an LLM,
e.g. a Llama \cite{touvron2023llama} model and then ask a question.
The three most relevant paragraphs from the project documents are retrieved.
The LLM is then prompted through the IBM Generative AI Python SDK \cite{ibm_generative_ai_sdk} to generate a summarised answer from these paragraphs, as well as individual answers from each paragraph.
We also return 
a Neo4j subgraph showing
all entity and relation objects from the three most relevant paragraphs.
We leave integrating the graph structure into the LLM prompt as future work.

\section{Use-case: Knowledge Discovery for the Battery Domain}

In this section, we show how KnowledgeHub can be used for an example project 
related to the battery domain.
The user starts by identifying and ingesting PDF documents related to their topic and creates their ontology based on BattINFO\footnote{\url{https://github.com/BIG-MAP/BattINFO}}.
The user annotates a document $d^1$ with 150 entity types on 1,988 spans.
They then train an NER model by fine-tuning \texttt{BatteryBERT-cased}\footnote{\url{https://huggingface.co/batterydata/batterybert-cased}} on $d^1$, and auto annotate a new document $d^2$ with 73 types on 1,464 spans.
This out-of-domain (OOD) auto annotation yields a micro F1 score of 54.8\% as shown in Table~\ref{tab:annotation-experiments}.
This eliminates the need for users to annotate unknown documents from scratch, reducing the cost of annotations by more than half. After revising the annotations of $d^2$, the user trains a new NER model on $d^1$ and $d^2$, and auto annotates another new document $d^3$ with 96 types on 1,467 spans. 
In-domain (ID) results are 52.8\%, 54.9\% and 61.9\% on $d^1$, $d^{1,2}$ and $d^{1,2,3}$, respectively, showing that repeating this process increases the performance of the model.

\begin{figure}
    \centering
    \includegraphics[width=0.44\textwidth]{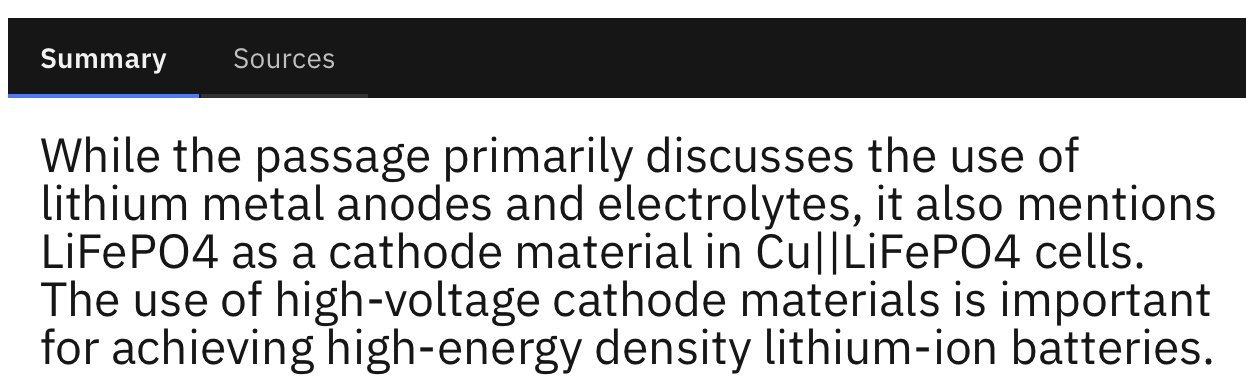} 
    \caption{Part of summary of the model answers over retrieved contexts.}
    \label{fig:sample-qa}
\end{figure}

\begin{figure}
    \centering
    \includegraphics[width=0.48\textwidth]{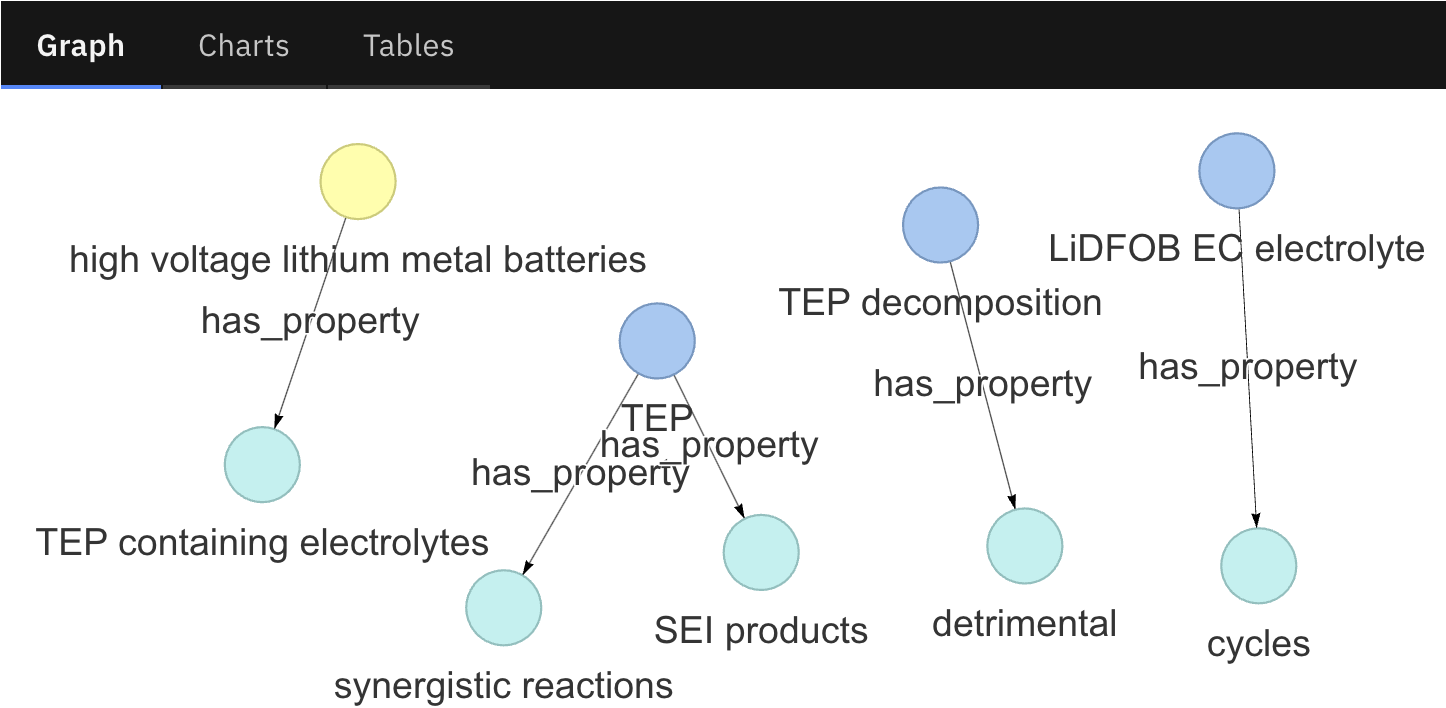} 
    \caption{Part of the entities/relations contained in the retrieved contexts.}
    \label{fig:sample-kg}
\end{figure}

The user can also perform QA over their documents,
where we show the summary using
the instruction-tuned Mistral model\footnote{\url{https://huggingface.co/mistralai/Mistral-7B-Instruct-v0.2}}
for an example question,
\emph{``What are promising cathode materials for high-voltage lithium-ion batteries?''} in Figure~\ref{fig:sample-qa}.
The KG showing the entities and relations in the retrieved contexts is shown in Figure~\ref{fig:sample-kg}.

\begin{table}[ht]
\centering
\small
\begin{tabular}{l|c|c|c}
\toprule
\textbf{Setting} & \textbf{Train} & \textbf{Dev} & \textbf{F1} \\
\midrule
OOD & $d^1$ & $d^2$ & 54.8 \\
ID & $d^1_{train}$ & $d^1_{dev}$ & 52.8 \\
ID & $d^1_{train}, d^2_{train}$ & $d^1_{dev},d^2_{dev}$ & 54.9 \\
ID & $d^1_{train}, \dots, d^3_{train}$ & $d^1_{dev}, \dots d^3_{dev}$ & 61.9 \\

\bottomrule
\end{tabular}
\caption{Performance of the auto annotation in Out Of Domain (OOD) and In Domain (ID) settings on documents $d^1$, $d^2$ and $d^3$.}
\label{tab:annotation-experiments}
\end{table}


\section{Conclusion}
In this paper we have presented KnowledgeHub,
a tool for assisted scientific discovery
by supporting IE tasks such as
NER and RC.
We also include a KG and an RAG component for grounded summarisation and QA,
enhancing the factual correctness of LLM responses.
We have demonstrated the usefulness of KnowledgeHub
through an example where a researcher uses the tool
for assisting their research for a project relating to batteries.

In future work,
we would like to 
explore more ways of combining the graph information and the retrieved contexts.
We would also like to implement QA based on non-text items in the PDF such as tables and figures.
We wish to support direct annotation on the PDF content
and improve functionality to support inter-annotator agreement.


\appendix

\section*{Ethical Statement}
We did not identify any ethical issues.

\section*{Acknowledgments}

This work was supported by the Hartree National Centre for Digital Innovation (HNCDI), a collaboration between STFC and IBM.


\bibliographystyle{named}
\bibliography{ijcai24}

\end{document}